\documentclass{aastex}
\usepackage{natbib}
\usepackage{emulateapj5}
\slugcomment{To appear in {\rm Publications of the Astronomical Society of the Pacific}, January 2002}
\shorttitle{Berkeley Extreme and Far-UV Spectrometer}
\shortauthors{Dixon, Dupuis, \& Hurwitz}

\begin{document}
 

\newcommand{\cf}{cf.}
\newcommand{\eg}{e.g.}
\newcommand{\etal}{et~al.}
\newcommand{\etc}{etc.}
\newcommand{\fig}[1]{Fig.~\ref{#1}}
\newcommand{\h}{$^{\rm h}$}
\newcommand{\ie}{i.e.}
\newcommand{\m}{$^{\rm m}$}

\newcommand{\cfour}{\ion{C}{4}}
\newcommand{\ebv}{$E($\bv)}
\newcommand{\f}{{\it f}\/}
\newcommand{\hone}{\ion{H}{1}}
\newcommand{\htwo}{H$_2$}
\newcommand{\kms}{km s$^{-1}$}
\newcommand{\nho}{$N$(\hone)}
\newcommand{\nht}{$N$(H$_2$)}
\newcommand{\oone}{\ion{O}{1}}
\newcommand{\osix}{\ion{O}{6}}
\newcommand{\rv}{$R_V$}
\newcommand{\specfit}{{\small SPECFIT}}

\newcommand{\chips}{{\it CHIPS}}
\newcommand{\cop}{{\it Copernicus}}
\newcommand{\euve}{{\it EUVE}}
\newcommand{\fuse}{{\it FUSE}}
\newcommand{\hst}{{\it HST}}
\newcommand{\hut}{HUT}
\newcommand{\iue}{{\it IUE}}
\newcommand{\orf}{{ORFEUS}}
\newcommand{\rosat}{{\it ROSAT}}

\newcommand{\na}{New~Astronomy}

\title{Final Calibration of the Berkeley Extreme and Far-Ultraviolet
Spectrometer on the ORFEUS-SPAS I and II Missions\footnotemark}

\footnotetext[1]{Based on the development and utilization of ORFEUS
(Orbiting and Retrievable Far and Extreme Ultraviolet
Spectrometers), a collaboration of the Institute for Astronomy and
Astrophysics at the University of T\"{u}bingen, the Space
Astrophysics Group of the University of California, Berkeley, and
the Landessternwarte Heidelberg.}

\author{W. Van Dyke Dixon\altaffilmark{2}, Jean Dupuis\altaffilmark{2},
and Mark Hurwitz}
\affil{Space Sciences Laboratory\\
University of California, Berkeley, Berkeley, California 94720-7450}
\email{vand@ssl.berkeley.edu, jdupuis@ssl.berkeley.edu, markh@ssl.berkeley.edu}

\altaffiltext{2}{Present address: Department of Physics and Astronomy,
The Johns Hopkins University, Baltimore, MD 21218}

\begin{abstract}

The Berkeley Extreme and Far-Ultraviolet Spectrometer (BEFS) flew as
part of the ORFEUS telescope on the {\it ORFEUS-SPAS~I\/ {\rm and}
II}\/ space-shuttle missions in 1993 and 1996, respectively.  The data
obtained by this instrument have now entered the public domain.  To
facilitate their use by the astronomical community, we have
re-extracted and re-calibrated both data sets, converted them into a
standard (FITS) format, and placed them in the Multimission Archive at
Space Telescope (MAST).  Our final calibration yields improved
wavelength scales and effective-area curves for both data sets.

\end{abstract}
 

\section{INTRODUCTION}

The Berkeley Extreme and Far-Ultraviolet Spectrometer (BEFS) flew
aboard the \orf\ telescope on the {\it ORFEUS-SPAS~I\/ {\rm and} II}\/
space-shuttle missions in 1993 September and 1996 November/December,
respectively.  The instrument returned 109 observations of 75
astrophysical objects on the first mission and 223 observations of 108
astrophysical objects on the second.
We have performed a new extraction and calibration of the BEFS spectra
from both missions.  The resulting data set offers an
improved flux calibration and wavelength solution in a standard FITS
data format.  The extracted spectra, together with the photon-event lists
from which they were derived, are available through the archives of
the Space Telescope
Science Institute.  Now in the public domain, these data will
contribute to studies of the interstellar medium; early-type, coronal,
and white-dwarf stars; cataclysmic variables; active galactic nuclei;
and the emission processes of the earth's outer atmosphere.  In
addition, they will be useful to investigators planning observations
with the {\it Far Ultraviolet Spectroscopic Explorer (FUSE)}.

In this paper, we discuss the final calibration of the Berkeley
spectrometer.  We describe the format of the resulting data files and
the software tools provided for their analysis.

\begin{table*}[t]
\vskip 10pt
\label{gratings}
\begin{center}
\small
{\sc Table 1\\
Berkeley Extreme and Far-Ultraviolet Spectrometer Gratings}
\vskip 2pt
\begin{tabular}{ccccccc}
\tableline
\tableline
& Central Groove & \multicolumn{2}{c}{Bandpass (\AA)} & \multicolumn{1}{c}{   } &
\multicolumn{2}{c}{Coating} \\ \cline{3-4} \cline{6-7}
Grating &
Density (mm$^{-1}$) &
1993 & 1996 & & 1993 & 1996 \\
\tableline
A & 6000.0 & 380--510  & 390--520  & & Evaporated Ir & Evaporated Ir \\
B & 4550.0 & 514--687  & 500--670  & & Evaporated Ir & Evaporated Ir \\
C & 3450.4 & 658--894  & 680--910  & & Evaporated Ir & Sputtered SiC \\
D & 2616.6 & 865--1175 & 910--1220 & & Evaporated Ir & Sputtered SiC \\
\tableline
\end{tabular}
\vskip 2pt
\parbox{5.0in}{
\small\baselineskip 9pt
\indent
{\sc Note.}---The geometry of each grating/detector system is
identical save for rotation about the Z axis and/or reflection through
the detector midplane.  The incidence angle is 12\arcdeg.  The
wavelength coverage ($\lambda_{\rm MAX} / \lambda_{\rm MIN}$) of each
grating is identical, as is the ratio of central wavelength to groove
density.}
\end{center}
\end{table*}

\section{DATA EXTRACTION AND CALIBRATION}

\subsection{The Instrument}

The BEFS sits at the prime focus of the 1-m \orf\ telescope
\citep{Grewing:91} To improve the resolution and reduce astigmatism,
the spectrometer employs mechanically-ruled diffraction gratings with
varied line spacing in a non-Rowland mounting.  Four such gratings,
together with an optic that feeds a fine-guidance sensor, are arranged
in an annulus at the end of the spectrometer opposite the entrance
aperture as shown in \fig{schematic}.  Each grating disperses a unique
spectral region (Table \ref{gratings}) onto one of two
microchannel-plate detectors located near the focal plane of the
spectrometer.  Together, the four gratings simultaneously span the
wavelength region from approximately 380 to 1220 \AA\ at a spectral
resolution of $\lambda/\Delta\lambda \sim 4600$ for point sources.  A
detailed discussion of the instrument's design may be found in
\citet{Hurwitz:86, Hurwitz:IAU:96}.  Its performance on the {\it
ORFEUS-SPAS~I}\/ mission is described in \citet{Hurwitz:95} and on the
{\it ORFEUS-SPAS~II}\/ mission in \citet{ORFEUS98}.

Spectra A and B are imaged by Detector 0, spectra C and D by Detector 1
(\fig{image}).  Between the 1993 and 1996 flights, the four gratings
were shifted in their mountings to optimize their wavelength coverage.
On the second flight, both spectra A and B span the important
\ion{He}{1} edge at 504 \AA, and spectrum D extends to wavelengths just
longer than the Lyman $\alpha$ line at 1216 \AA.  On the 1993 flight,
both detectors were powered for nearly all observations, but on the
1996 flight, Detector 0 was powered only during observations of known
EUV sources in order to minimize its contribution (mostly in the form
of detector background events) to the telemetry stream.  For both
missions, Grating A and B spectra were extracted only for selected
targets.

An unplanned change in the spectrograph for the 1996 flight was an
apparent shift of one or both of Gratings A and B; their spectra
intersect, forming an ``X'' on the detector (\fig{image}).  The most
likely reason for the grating shift is ``creep'' in the viton material
that was employed to constrain the optics within their holders.
Because these two optics were not de-mounted between flights, the
viton, which is normally under a small shear load, may have crept or
relaxed, leaving the grating susceptible to shifts under the stress of
launch.  Gratings C and D show no such shift, perhaps because they were
de-mounted after the first mission for overcoating with SiC.

On the 1993 flight, a single aperture 20\arcsec\ in diameter was used
for all astronomical observations.  On the 1996 flight, the
spectrograph was equipped with a diaphragm selection blade with three
scientifically-useful positions:  Position 1 contains a single on-axis
aperture 20\arcsec\ in diameter; it was rarely used for science
observations.  Position 2 contains a single on-axis aperture
10\arcmin\ in diameter, used during the initial star tracker/telescope
co-alignment and for some observations of diffuse backgrounds and
extended objects.  Position 3, with which most science observations
were obtained, contains three apertures.  Targets were usually placed
in the on-axis aperture, which is 26\arcsec\ in diameter.  A second
clear aperture, about 1.4 times larger in area and displaced by
2\farcm4, was designed to obtain simultaneous airglow observations,
although in crowded fields an astrophysical source was occasionally
observed.  A third aperture, 120\arcsec\ in diameter, was displaced by
5\farcm0 and covered by a tin filter approximately 1500 \AA\ thick.
The tin filter is virtually opaque to FUV radiation and was employed
primarily for observations of the EUV spectrum of the bright B-type
star $\epsilon$~CMa.

In our nomenclature, on-axis (target) spectra are designated as type 1,
off-axis (airglow) spectra as type 2, and spectra obtained through the
tin filter as type 3.  While the {\it ORFEUS-SPAS~I}\/ data set
consists only of spectra A1, B1, C1, and D1, up to twelve separate
spectra may be extracted from the {\it ORFEUS-SPAS~II}\/ data, three
from each grating.  In practice, however, we extract the airglow
spectrum only from Grating D and the tin-filter spectrum only from
Grating B, reducing the actual number of spectra to six: A1, B1, B3,
C1, D1, and D2.  Airglow spectra from Gratings A, B, and C are less
useful than those from Grating D:  the Grating A bandpass contains no
geocoronal features, and spectrum B2 is contaminated by flux from
spectrum A1.  For most targets, spectrum C1 contains no astrophysical
flux, so extracting an airglow spectrum is redundant.  Complete photon
lists for each detector, along with tools for extracting spectra from
them, are available from the archive (see Section \ref{format}).

\subsection{Spectral Extraction}
\label{extraction}

Each photon event processed by our electronics requires 24 bits of
encoding: 15 bits for the X (dispersion direction) coordinate, 8 bits
for Y, and 1 bit for detector identification.  The image size is thus
32,768 $\times$ 256 pixels.  In the FUV (900--1200 \AA), a single pixel
in X corresponds to a wavelength interval of approximately 0.014 \AA.
The instrumental resolution is $\sim$ 0.23 \AA, so the spectra are
highly oversampled in wavelength.  In the spatial domain, one X pixel
corresponds to about 0\farcs3 of sky, one Y pixel to about 10\arcsec.

For the {\it ORFEUS-SPAS~I}\/ data set, we find that, at a given
detector X coordinate, the FWHM of our spectra averages 5 pixels in the
Y dimension.  We thus sum the counts over 11 Y pixels (best-fit center
$\pm5$ pixels), a range that encompasses more than 98\% of the
dispersed photon events.  The general background is scaled from two
strips, each spanning 3 pixels in Y, immediately above and below the
spectral region of the detector.  For the second mission, improvements
in the detector electronics tightened the point-spread function in both
dimensions.  Accordingly, we use a 9-pixel window, with three pixels of
background on either side, in our extraction of spectra obtained on
{\it ORFEUS-SPAS~II}.

Extraction windows were defined once for each spectrum and each
flight.  The spectra are not observed to drift out of the standard
extraction windows during the {\it ORFEUS-SPAS~I}\/ mission, but move
slightly during the first few days of the {\it ORFEUS-SPAS~II}\/
flight.  We correct for these motions by shifting the standard
extraction window by up to three pixels in Y.  The motions are largest
in the EUV spectra.  Grating D spectra do not show significant motion
except for one observing period (GMT day 330) during the {\it
ORFEUS-SPAS~II}\/ mission.  On this flight, the spacecraft was twice
required to assume a ``minimum drag'' configuration, in which the
platform orientation was held fixed relative to the ram vector.  After
the second minimum-drag episode, which lasted for 12 hours, thermal
stresses in the spacecraft resulted in a significant shift in the
position of the spectrum on the detector, requiring a re-definition of
the extraction window for observations in this time frame.  The data
quality does not seem to have been compromised.  The minimum-drag event
and its consequences are discussed in \citet{ORFEUS98}.

For each observation, a photon-event list is extracted from the
spacecraft data stream.  A preliminary extraction of the FUV spectrum
(if available and sufficiently bright) is used to produce a plot of
count rate versus integration time.  These plots are manually
inspected, and ``good times'' (target in aperture, background rate low,
telemetry uninterrupted, etc.) are identified.  Status information from
the spacecraft star tracker is used to establish good times when
spectrograph data cannot.  Only data from these good times are included
in the photon-event lists and extracted spectra that
are supplied to the data archives.  Users interested in time-series
analyses should determine whether time segments have been excluded from
the observation of interest.

\subsection{Dead Time and Gain Sag}
\label{gain_sag}

When the spectral photon-event rate approaches the capacity of the
telemetry system (about 4400 events s$^{-1}$), instrument dead-time
effects become important.  The detector electronics create ``stimulation''
events (or stim pins) at the detector edges at a rate ($\sim 2$ s$^{-1}$) that can be
measured during slews and other quiet periods.  The observed rate of these
events during on-target pointings enables us to estimate the dead time
for genuine spectral photons.  Data buffering and the non-periodic nature
of the telemetry-sampling system ensure that the throughput for both
stimulation and actual photon events is identical.

When the current flow parallel to the microchannel plate surface
cannot replenish the electrons being drained from the channel walls,
the modal gain may sag. Gain sag is a function of local count rate
caused by unusually bright emission-line or continuum targets and
must therefore be distinguished from overall dead-time effects, which
are a function of total count rate, including background. Gain sag is
of greatest concern near the center of the band, where the astigmatism
or spectral height is minimized.  In the Y-dimension, gain sag causes a
loss of spatial resolution, with the result that some source events are
not contained within the nominal extraction window. (Spectral spillover
unrelated to gain sag may also occur if the source falls very near the
edge of the entrance aperture.) The remedy is to study the extracted
background spectra for evidence of spectral spillover and to adjust the
background-subtraction procedure accordingly.

The delay-line system that calculates the photon X coordinate is
generally robust against gain sag, but high local count rates can lead
to wavelength-dependent variations in throughput.  A very bright
continuum source may suffer an uncorrected loss of throughput over
broad regions in which the astigmatism is small.  We find a depression
of the observed continuum whenever the local flux exceeds approximately
$5 \times 10^{-11}$ erg cm$^{-2}$ s$^{-1}$ \AA$^{-1}$ for wavelengths
less than about 1110 \AA.  At longer wavelengths, instrumental
astigmatism broadens the spectrum in Y, reducing the local count rate.
In addition, spurious absorption features can be observed at
approximately 1047, 1052.5, 1057, and 1060 \AA.  We have determined
that these features are fixed in detector coordinates, but may
vary somewhat in their assigned wavelengths.  For each spectrum that
exceeds this brightness limit, warnings have been placed in the
corresponding file headers.

\subsection{Pointing Effects}
\label{pointing}

For fine pointing, the spacecraft's attitude control system (ACS)
utilizes signals from a star tracker aligned with the \orf\ telescope
axis.  Under normal circumstances, the ACS achieves an absolute
pointing error of less than $\sim 5$\arcsec.  For a few targets, plots
of count rate versus time clearly indicate that the target was not
centered in the aperture but fell near the edge; the spectra obtained
during these pointings are labeled (in both the file headers and
published catalogs) as non-photometric.  Pointing jitter was about $\pm
2$\arcsec\ peak to peak on both flights of the \orf\ telescope.  No
attempt has been made to correct our spectra for the motion of the
target in the aperture.

Because the telescope and star tracker were aligned using BEFS Grating
D spectra, any defocus in the telescope (relative to the aperture plane
of the spectrograph) would cause light destined for the other gratings
to be displaced systematically toward the aperture edge.  We believe
that some spectra obtained with Gratings A and B on the {\it
ORFEUS-SPAS~II}\/ mission suffer partial occultation, both because
variations in the measured flux are greater than we can otherwise
account for, and because sources observed with Grating B through the
much larger tin-filter aperture do not show flux variations.  Because
the occultation effects cannot be calculated {\it a priori}, BEFS EUV
spectra must be treated as non-photometric.

\subsection{Backgrounds}
\label{backgrounds}

The on-orbit background event rate, measured on an unilluminated region
of the FUV detector during a night-time observation, is about 0.0016
counts per second per resolution element (9 pixels in Y by 25 in X),
due mostly to detector dark counts and cosmic rays.  The rate measured
in an extracted spectrum far from any airglow feature ranges from
about 0.003 s$^{-1}$ at night to 0.04 s$^{-1}$ in some daytime
pointings, probably due to stray Lyman $\alpha$ radiation.  (The
daytime rate of 0.0044 s$^{-1}$ quoted in \citealt{ORFEUS98} is a
typographical error.)

The day and night count rates in the ``bright corner,'' a region of
enhanced background flux present on both detectors on both missions (see \fig{image}),
are approximately 0.09 and 0.01 s$^{-1}$, respectively.  Their ratio
corresponds roughly to the day/night intensity ratio of diffuse Lyman
$\alpha$ emission.  In the 1993 data set, the bright corner contaminates Grating A
spectra at wavelengths shorter than about 420 \AA\ and Grating D
spectra shorter than about 960 \AA.  In the 1996 data set, Grating D
spectra are contaminated at wavelengths below about 1000 \AA, while the
accidental shift of Grating A moved its spectrum away from the bright
corner.

After the {\it ORFEUS-SPAS~I}\/ mission, substantial baffling was added
to the ORFEUS Echelle Spectrometer \citep{Barnstedt:99} with
which the telescope was shared.  It was expected that this baffling
would eliminate the bright corners on the Berkeley detectors.  The
effect, however, is essentially unchanged in the {\it ORFEUS-SPAS~II}\/
data set and has since been attributed to a zero-order reflection from
the Berkeley gratings of a light leak at a hardware interface seam.

\begin{figure}
\plotone{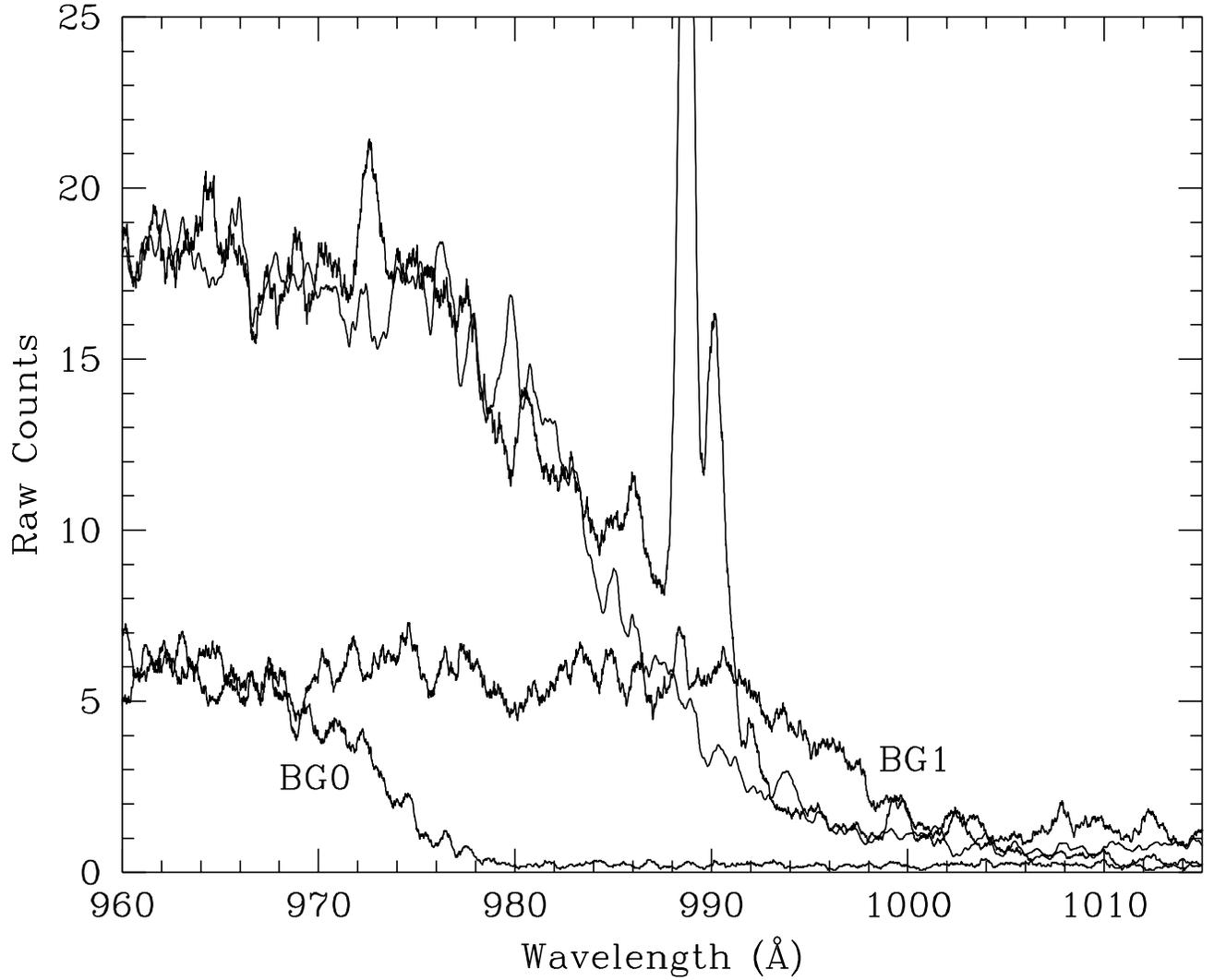}
\figcaption{The edge of the bright corner seen in an {\it ORFEUS-SPAS~II}\/ airglow
spectrum and its associated background spectra.  In this region,
the background cannot simply be scaled from the sum of
the 3-pixel background strips above (BG1) and below (BG0) the target
spectrum.  Instead, we shift both spectra by 600 pixels in X ($\sim$
8.6 \AA), then sum and scale them to the width of the extraction
window.  In this figure, all spectra have been box-car smoothed by 41
pixels ($\sim$ 0.6 \AA) for clarity.   The emission feature is the
\oone\ complex near 989 \AA.
\label{bright_corner}
}
\end{figure}

The irregular footprint of the bright corner creates a crossover region
in our FUV spectra (Grating D) within which the background cannot
simply be scaled from the 3-pixel strips above and below the spectrum.
To estimate the background here, we assume that the shape of the
background in the spectral region and in the adjacent strips differs
only by a translation in the X coordinate, determined from detector
images collected during slew periods.  The process is illustrated in
\fig{bright_corner}, which shows part of an airglow spectrum obtained
on the {\it ORFEUS-SPAS~II}\/ mission.  To model the background between
965 and and 1022 \AA, we shift the background spectra extracted from
the regions above (BG1) and below (BG0) the target spectrum by 600
pixels in X ($\sim$ 8.6 \AA; BG1 is shifted to the left, BG0 to the
right), then scale their sum to the width of the target extraction
window.  For the {\it ORFEUS-SPAS~I}\/ dataset, we shift the background
strips by $\pm 750$ pixels in X ($\sim$ 10.7 \AA) in the region between
920 and 977 \AA.  It is possible for this procedure to introduce
artificial features within the overlap region.  Artifacts of this type
are not expected to be as narrow as unresolved spectral features, nor
are they likely to exceed 10\% of the background in amplitude. As such,
they are of potential concern only for very faint targets.

The extraction of Grating A spectra is complicated by two effects.  On
the 1993 flight, wavelengths shorter than about 420 \AA\ are
contaminated by the bright corner.  The background in the crossover
region (about 390--420 \AA) is estimated as for the Grating D spectra,
and the same caveats apply.  On the 1996 flight, the Grating A spectra
moved away from the bright corner, but were left crossing the Grating B
spectra, contaminating spectrum A1 between 400 and 460 \AA\ and
spectrum B1 between 550 and 590 \AA.  We have developed an algorithm to
estimate the contamination of spectrum A1 by shifting and adding the
background strips (similar to the scheme used for the edge of the
bright corner), but the resulting background array provides only a
rough estimate of the contamination due to spectrum B1.  Researchers
interested in absorption features or other spectral details should
consult the raw-counts spectra, which are not background subtracted, to
confirm the reality of apparent spectral features in this band.  The
higher effective area of Grating B (Section \ref{orf2}) makes the
contribution of the overlap to spectrum B much less significant, and we
do not attempt to correct for it.

\subsection{Spectral Resolution and Wavelength Scale}

The BEFS monochromatic point-spread function (PSF) is determined by
fitting Gaussian line profiles to emission features in the spectrum of
the bright symbiotic binary \objectname{RR Tel}.  \citet{Penston:1983}
list the mean line width for each emitting species, determined from
high-resolution \iue\ spectra.  Assuming that the intrinsic line width
and instrument resolution add in quadrature, \citet{ORFEUS98} derive
independent PSF values from the lines of eight emitting species and
quote a mean instrumental resolution of $\sim 95$ \kms\ for the BEFS
spectrograph.  We have repeated this analysis using our re-extracted
spectra and the updated line list presented in Table 2.
(Line identifications were kindly provided by Brian Espey.)  Emission
lines from nine species listed in Penston \etal\ are present in the
BEFS spectrum of RR~Tel.  Their intrinsic (\iue) widths range from 43
to 78 \kms\ (FWHM), and their measured (BEFS) widths from 79 to 187
\kms\ (FWHM).  The resulting PSF values range from 63 to 175 \kms.
While our results are similar to those of Hurwitz \etal, we note that
six of the nine species yield PSF values between 60 and 70 \kms.
RR~Tel is a dynamic system, and the three high values, from
\ion{He}{2}, \ion{C}{4}, and \ion{O}{4}, may well reflect changes in
the intrinsic widths of these lines over the decade between the
\iue\ and BEFS observations.  Rejecting the three high values, we
average the remaining six to find a mean instrumental PSF of
\linebreak[4]

\begin{figure}
\plotone{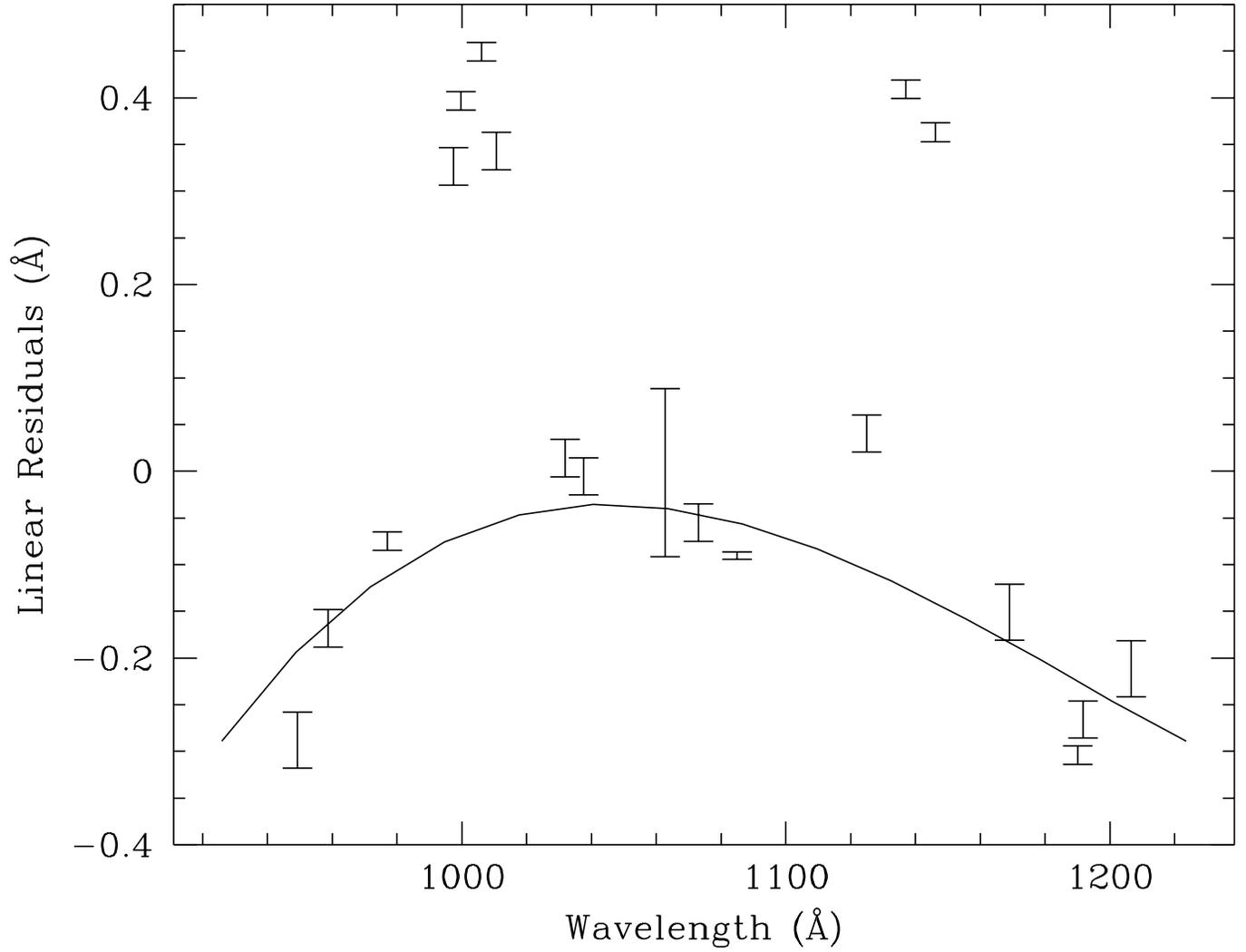}
\figcaption{The BEFS wavelength calibration.
The points mark the measured positions of
emission features in the spectrum of RR Tel, plotted as 
residuals from a linear fit.  The solid curve represents the
deviation from a linear wavelength solution (wavelength minus linear
fit) derived from an optical ray trace of the telescope and
spectrometer.  The solid curve has been shifted in both
dimensions to reproduce the observed RR Tel wavelengths, excluding the
sharp deviations around 1000 and 1140 \AA.  Additional wavelength
points, derived from the spectra of white dwarfs and $\alpha$ CMi, 
are not shown.
\label{wavecal}
}
\end{figure}

\vskip -20pt
\vbox to 3.2in {
\begin{center}
\small
{\sc Table 2\\
Emission Features in the Spectrum of RR Tel}
\vskip 2pt
\begin{tabular}{ccccc}
\tableline
\tableline
& $\lambda_{lab}$ & \phm{XXXX}&& $\lambda_{lab}$ \\
Feature & (\AA)   && Feature & (\AA) \\
\tableline
\ion{S}{6}   & \phn933.38  && \ion{O}{4}   &    1067.81 \\
\ion{He}{2}  & \phn942.51  && \ion{S}{4}   &    1072.97 \\
\ion{S}{6}   & \phn944.52  && \ion{N}{2}   &    1083.99 \\
\ion{He}{2}  & \phn949.33  && \ion{He}{2}  &    1084.94 \\
\ion{He}{2}  & \phn958.70  && \ion{P}{2}   &    1124.95 \\
\ion{C}{3}   & \phn977.02  && \ion{Ne}{5}] &    1136.66 \\
\ion{Ne}{6}] & \phn997.40  && \ion{Ne}{5}] &    1145.83 \\
\ion{Ne}{6}] & \phn999.63  && \ion{P}{2}   &    1152.83 \\
\ion{Ne}{6}] &    1006.10  && \ion{C}{4}   &    1168.94 \\
\ion{Ne}{6}] &    1010.60  && \ion{Mg}{6}] &    1190.07 \\
\ion{O}{6}   &    1031.93  && \ion{Mg}{6}] &    1191.64 \\
\ion{O}{6}   &    1037.62  && \ion{S}{5}   &    1199.18 \\
\ion{S}{4}   &    1062.66  && \ion{Si}{3}  &    1206.51 \\
\tableline
\end{tabular}
\vskip 2pt
\parbox{3.5in}{
\small\baselineskip 9pt
\indent
{\sc Note.}--- This set of lines differs slightly from that used to define
the BEFS wavelength scale (shown in \fig{wavecal}); it has been revised
to reflect recent results.}
\end{center}
}
\addtocounter{table}{1}

\noindent
$\sim 66$ km s$^{-1}$, or 0.23 \AA.  This result is consistent with both
preflight laboratory measurements of the BEFS resolution
\citep{ORFEUS98} and recent analyses of interstellar molecular-hydrogen
absorption-line spectra \citep{Dixon:Hurwitz:Lee:01}.

Over most of the FUV bandpass, the relationship between the observed
wavelength and the detector X coordinate is well modeled by a linear
function of X plus a smooth departure that does not exceed $\pm 0.13$
\AA.  This departure from linearity, derived from an optical ray trace
of the telescope and spectrograph, is presented in \fig{wavecal}.  The
ray trace provides the shape of this curve, but not its absolute
position in wavelength space.  We have thus shifted the curve to best
reproduce the measured positions of emission lines in the spectrum of
RR~Tel, also plotted in \fig{wavecal}.  In two wavelength regions, 990--1010
and 1125--1145 \AA, the line positions in the RR Tel spectrum show
significant deviations from our ray-trace predictions.  In the end, we
decided not to model these deviations, but instead exclude these
regions when fitting our model curve to the RR Tel line positions.  We
supplement the wavelength solution at short wavelengths with
Lyman-series absorption lines in several white-dwarf stars and at long
wavelengths with emission features from the late-type star $\alpha$
CMi.

Temporal changes in the FUV wavelength scale, seen on both flights, are
modeled with a time-dependent offset and a linear stretch.  We combine
repeated observations of diffuse emission (airglow) lines and RR Tel to
characterize these effects.  (RR Tel was observed only once on the {\it
ORFEUS-SPAS~I}\/ mission.) This new wavelength calibration represents a
significant improvement over those initially provided with the {\it
ORFEUS-SPAS~I}\/ data set, which suffered from significant
non-linearities, and the {\it ORFEUS-SPAS~II} spectra, which were
poorly constrained at the longest wavelengths.  To establish a
wavelength scale for the off-axis aperture, we apply a fixed offset,
determined from the average of the strongest diffuse emission features,
to the on-axis wavelength solution.

The EUV wavelength scale is derived from pre-flight laboratory
measurements.  For spectra B and C, the airglow features \ion{He}{1}
$\lambda 584.33$ and \ion{O}{2} $\lambda \lambda 832.8$--834.5
(respectively) provide a fixed point for the wavelength calibration and
a measure of its temporal drift.  Because spectrum A contains no
measurable emission features, its wavelength calibration is
considerably more uncertain.

The wavelengths assigned by our spectral-extraction software reflect
the effects mentioned above, as well as spectral shifts due to
anomalous thermal conditions following the second minimum-drag
period on the {\it ORFEUS-SPAS~II}\/ mission.  Relative wavelength
errors within a given spectrum should be less than $\sim 0.2$
\AA\ between 915 and 1216 \AA.  (The {\it ORFEUS-SPAS~I}\/ spectra
extend only to 1175 \AA.)  Because of the unknown placement of each
target within the aperture, an overall offset of $\pm 0.5$ \AA\ may
exist in any individual FUV spectrum.

\subsection{Flux Calibration}
\label{flux_cal}

Flux calibration of the BEFS is based on in-flight observations of hot
DA white dwarfs.  The efficacy of this method for on-orbit calibration
in the FUV has been demonstrated by \citet{HUT1CAL2,HUT2CAL2}.  Our
principal calibration stars are \objectname{G~191--B2B} for the {\it
ORFEUS-SPAS~I}\/ data set and \objectname{HZ~43} for {\it ORFEUS-SPAS
II}.  Given the uncertainties in the model atmospheres and synthetic
spectra, the adopted stellar {\it V}\/ magnitudes and temperatures, and
the photometric accuracy and background subtraction of the extracted
count-rate spectra, we quote an uncertainty in the BEFS flux
calibration of $\lesssim$ 10\%.

\subsubsection{{\it ORFEUS-SPAS~I}}

\citet{Vennes:96} model the FUV spectrum of G~191--B2B obtained on the
{\it ORFEUS-SPAS~I}\/ mission.  Following them, we use the
pure-hydrogen models of \citet{Vennes:92} and adopt the stellar
parameters $T_{\rm eff} = 60,000$ K, $\log g = 7.535$, and $V =
11.781$.  $T_{\rm eff}$ and $\log g$ represent an average of the
maximum and minimum values found in the literature \citep{Bergeron:94,
Finley:97, Marsh:97, Vennes:97} and are in excellent agreement with
those used by \citet{Bohlin:95} and \citet{HUT1CAL2} to flux-calibrate
the {\it Hubble Space Telescope} / Faint Object Spectrograph (FOS) and
Hopkins Ultraviolet Telescope (\hut), respectively.  The {\it V}\/
magnitude is taken from the compilation of \citet{Bohlin:95}.  We scale
the model by the \hone\ interstellar opacity, adopting a column density
of $\log N($\hone) $= 18.31 \pm 0.01$ \citep{Dupuis:95}, a value
consistent with the recent \hst\ result of \citet{Vidal-Madjar:98}, and
add a synthetic second-order spectrum scaled to the observed flux
between 865 and 912 \AA.  (The ratio of second- to first-order counts
in the BEFS FUV spectrum of G~191--B2B is $\sim$ 0.02.) The model is
then convolved with the instrument-resolution profile.  An
effective-area curve is derived by dividing the count-rate spectrum of
G~191--B2B by the final synthetic stellar spectrum.  The result is
fitted with a fifth-order Legendre polynomial to remove structure on
small spatial scales.

G~191--B2B was observed four times on the {\it ORFEUS-SPAS~I}\/
mission.  We use three other stars, MCT 0501--2858, MCT 0455--2812, and
HD~220172, as transfer standards to establish effective-area curves
for the first and final days of the mission.  The resulting family of
effective-area curves is shown in \fig{o1_flux}.  Effective-area curves
for Gratings A and B are derived from observations of G~191--B2B in the
same way as for Grating D.  The differences among the Grating A
effective-area curves appear to be real and indicate a decrease in
sensitivity with time, especially at long wavelengths.  The differences
among the Grating B curves, however, are not significant given the
signal-to-noise ratios of the individual spectra.  We thus average the
four curves shown into a single effective-area curve for this
spectrum.  Our Grating C spectrum of G~191--B2B exhibits no significant
first-order astrophysical flux; we adopt a theoretical effective-area
curve for this spectrum.

FUV spectra of hot stars with low \ion{H}{1} column densities are
contaminated by stellar EUV flux in second order.  For each such star,
we use spectra A and B to compute the second-order contribution to
spectra C and D, scaling a theoretical second-order effective-area
curve to match the raw counts observed in spectrum C and in spectrum D
below 912 \AA.  The resulting second-order spectra are provided for
users who wish to subtract the second-order contribution to their FUV
spectra.  Figure \ref{o1_flux} shows a number of second-order curves
for spectrum C.  The highest represents the effective area through the
first half of the 1993 flight; we apply an average of the remaining
curves to observations obtained in the second half.

\subsubsection{{\it ORFEUS-SPAS~II}}
\label{orf2}

The flux calibration for {\it ORFEUS-SPAS~II}\/ is based on
observations of HZ~43 as described in \citet{ORFEUS98}.  We have
repeated that calculation, but include a correction for the
second-order contribution to the observed FUV spectrum that was
neglected in the earlier analysis.  We use the white-dwarf models of
\citet{Vennes:92}, adopting the stellar parameters $T_{\rm eff} =
50,000$ K and $\log g = 8.0$.  These parameters were used by
\citet{Bohlin:95} for the flux calibration of the FOS and agree with
published values (e.g., \citealt{Napiwotzki:93}).  We normalize the
model to the star's visual magnitude ($V = 12.914$;
\citealt{Bohlin:95}) and scale by the interstellar hydrogen opacity,
assuming a column density of $\log N($\hone) $= 17.94 \pm 0.03$
\citep{Dupuis:95}.  We add the synthetic first- and second-order
spectra and convolve the result with the instrument-resolution profile
to obtain a model stellar spectrum.  To derive the first-order
effective-area curve, we divide the count-rate spectrum of HZ 43 by
this synthetic spectrum and fit the result with a fifth-order Legendre
polynomial to retain the overall shape of the curve while removing
structure on small spatial scales.

Because Grating D was shifted to longer wavelengths for the 1996
mission, it does not sample enough of the wavelength region shortward
of 912 \AA\ to constrain models of the second-order stellar spectrum.
Instead, we derive the instrument's second-order effective-area curve
using laboratory data for the mirror and grating reflectivity, detector
quantum efficiency, and grating efficiency, then scale the result
(slightly) to match the observed second-order \ion{He}{1} $\lambda 584$
geocoronal line.  A synthetic EUV spectrum of HZ~43 is multiplied by
the computed second-order effective area to model the star's
second-order contribution at FUV wavelengths.  We find that the ratio
of second- to first-order counts in the BEFS FUV spectrum of HZ~43
rises from 0.01 to 0.04 between 1000 and 1200 \AA.

\begin{figure}
\epsscale{0.8}
\plotone{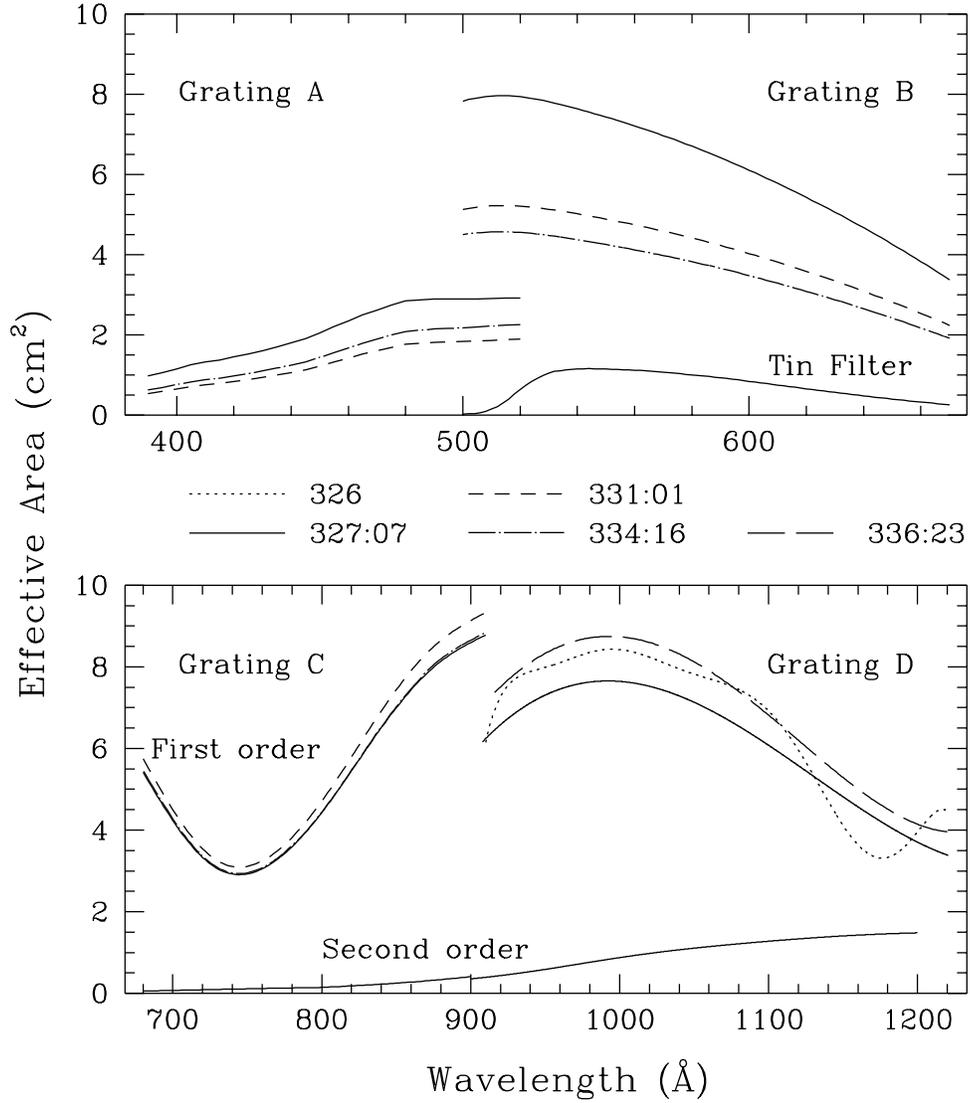}
\figcaption{Effective-area curves for the BEFS on the
{\it ORFEUS-SPAS~II}\/ mission, for which the principal
flux-calibration target is HZ~43.  The dramatic variation in the effective
areas of Gratings A and B reflects the photometric problems that affect
our observations at these wavelengths.  Rather than attempting to
correct for them, we adopt the first (highest) effective-area curve for
all Grating A and B spectra.  For Gratings C and D, we employ the curve
obtained nearest in time to each observation.  (We do not interpolate
among the curves.) The second-order curves are derived theoretically
and scaled to the observed second-order flux.
\label{o2_flux}
}
\end{figure}

The helium white dwarf MCT 0501--2858 is used as a transfer standard to
establish effective-area curves for the first and final days of the
mission.  Because we lack a reliable EUV model for this star, we use a
spectrum observed by the {\it Extreme Ultraviolet Explorer (EUVE)}\/ to
estimate its second-order flux.  The complete family of effective-area
curves is shown in \fig{o2_flux}.

In our assignment of effective-area curves to individual FUV spectra,
we interpolate linearly with time among the flux-calibration files for
the {\it ORFEUS-SPAS~I}\/ data set, as these vary smoothly through the
mission.  During the {\it ORFEUS-SPAS~II}\/ mission, the effective area
changes dramatically over the first two days of the mission, but
remains stable thereafter (\fig{o2_flux}).  We thus assign
specific effective-area curves to observations taken during each of the
first two observing shifts and use the mean of all subsequent curves
to calibrate spectra for the rest of the mission, as described in
\citet{ORFEUS98}.

On {\it ORFEUS-SPAS~II}, the effective area of the FUV channel falls
early in the mission, then rises and remains nearly
constant for the remainder of the 14-day flight.  Presumably, this
reflects condensation and evaporation of out-gassed material onto the
optics.  The effective area falls throughout the five-day {\it
ORFEUS-SPAS~I}\/ mission. The behavior differs between the two
missions, we believe, because the first flight was the maiden voyage of
the {\it ASTRO-SPAS}\/ spacecraft and thus subject to a higher level of
contamination.  (A second, two-week flight of {\it ASTRO-SPAS}, with
another payload, took place between the two {\it ORFEUS-SPAS}\/
missions.)

The long-wavelength dip in the effective area apparent early in both
missions reflects, we believe, variations in the ``pulse height'' or
amplitude of the charge cloud reaching the detector caused by
adsorption of molecules onto the microchannel plate (MCP) during
integration.  Subsequent ``scrubbing'' of the MCP during observations
on orbit desorbs the molecules, restoring the original MCP gain.  We
cannot test this hypothesis directly, because pulse-height information
is not included in the {\it ORFEUS-SPAS}\/ flight telemetry stream.
Preflight measurements, however, indicate higher-than-average gain in
the region of the detector that subsequently shows a dip in effective
area.  We therefore suspect a pulse-height effect.

Comparison of the {\it ORFEUS-SPAS~I}\/ and {\it II}\/ effective-area
curves reveals a dramatic increase in the spectrograph's FUV
sensitivity from the first flight to the second.  In preparation for
the second flight, the two long-wavelength diffraction gratings were
over-coated with silicon carbide, increasing their reflectivity at FUV
wavelengths.  The effective area of our Grating B spectra is
essentially unchanged from one flight to the next, but the Grating A
sensitivity is reduced by half.  For both channels, the effective area
falls precipitously for the first few days of the mission.  In fact, it
is likely that the sensitivity of our detector was fairly stable, but
that EUV observations of HZ~43, our flux-calibration target, suffered
the photometric problems described in Section \ref{pointing}.  Rather
than attempting to correct individual spectra for this effect (which
may differ greatly from one target to another), we have used a single
flux-calibration file, obtained early in the mission, for each of the
Grating A and B spectra obtained on the {\it ORFEUS-SPAS~II}\/
mission.  The absolute flux calibration of these spectra is therefore
suspect.

The tin filter, unavailable on the 1993 flight, was used in 1996 for
observations of the bright B-type star $\epsilon$ CMa and two
flux-calibration sources, HZ~43 and MCT~0501-2858.  The $\epsilon$ CMa
spectra were published \citep{Cohen:98} before the tin-filter
effective-area curve (plotted in \fig{o2_flux}) became available.
Observations obtained through the tin filter appear not to suffer the
flux losses seen in the other EUV data sets, perhaps because the
aperture containing this filter is much larger than the standard
observing aperture.

Because our primary flux-calibration star for the {\it ORFEUS-SPAS~I}\/
mission, G~191--B2B, shows no flux in the wavelength range covered by
spectrum C, we used a theoretical effective-area curve for the Grating
C spectra from that flight.  The {\it ORFEUS-SPAS~II}\/ calibration
target HZ~43 has a lower \hone\ column and exhibits enough flux at
short wavelengths (670--720 \AA) that we were able to scale the
theoretical effective-area curve to the observed intensity.

Second-order effective-area curves for Gratings C and D were derived
theoretically, as described above, then scaled to reproduce the
sub-Lyman-limit flux at the long-wavelength end of spectrum C and the
\ion{He}{1} $\lambda 584$ geocoronal line in spectrum D, respectively.
Because of the photometric problems apparent in our EUV data, it is
possible that an EUV-emitting source could contribute second-order flux
to a Grating D spectrum that is not accounted for in the
corresponding Grating A and B spectra.  Users should thus take care
when subtracting the second-order spectra provided with the {\it
ORFEUS-SPAS~II}\/ data.

\subsection{Flat-Field and Other Effects}
\label{flat_field_text}

\begin{figure}
\epsscale{1.0}
\plotone{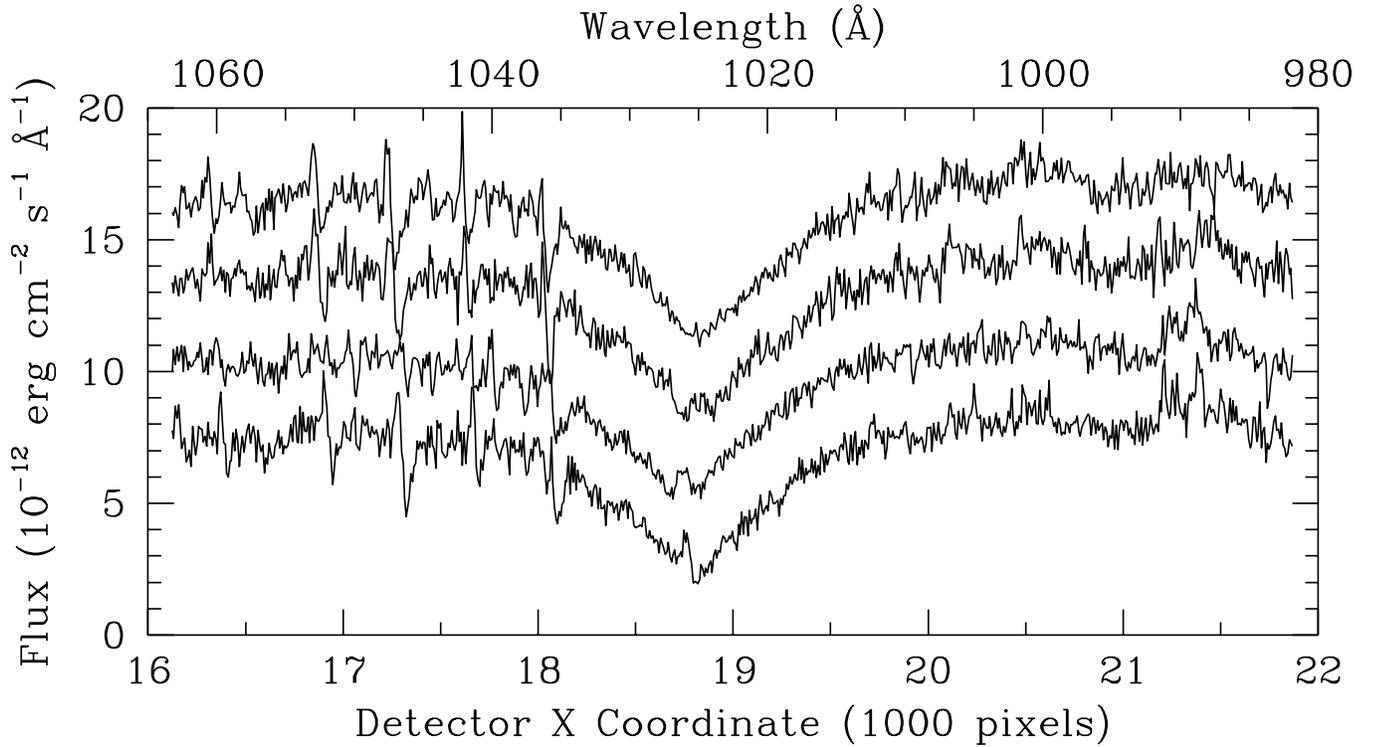}
\figcaption{Flat-field effects in {\it ORFEUS-SPAS~II}\/ spectra of
HZ~43.  Note the four P Cygni-like features between 1035 and 1055
\AA\ and the broad trough between 991 and 999 \AA.  These features are
discussed in the text.  From bottom to top, the spectra were obtained
on GMT days 327, 331, 334, and 336.  The flux of the bottom spectrum is
correct; the others are offset in flux space for clarity.  The
data are binned by 8 pixels and are plotted in pixel space; the
wavelength scale is approximate.
\label{flat_field}
}
\end{figure}

\citet{ORFEUS98} discuss the statistical distribution of fluctuations
in the response of the BEFS detector to uniform illumination.  We
examine two types of flat-field effects in \fig{flat_field}, which
presents four spectra of HZ~43 obtained over the course of the {\it
ORFEUS-SPAS~II}\/ mission.  The spectra are plotted in pixel space; the
wavelength scale is approximate.  Note the four narrow,
P Cygni-like features at about 1036, 1042, 1047, and 1053 \AA.
These features are nearly fixed on the detector, but vary somewhat in
wavelength space.  They are not seen in most BEFS spectra, which have
lower signal-to-noise ratios.  In much brighter targets, these are the
locations of the spurious absorption features discussed in Section
\ref{gain_sag}; the two effects are likely to be related.
At shorter wavelengths, \fig{flat_field} shows a broad trough in the
spectral continuum between about 990 and 1000 \AA.  Early in the
mission (bottom curve), the feature has sharp edges and a flat core;
later, the drop becomes more gentle and the feature considerably less
prominent.  We do not attempt to correct for flat-field effects in
the BEFS spectra.

The reader is referred to \citet{ORFEUS98} for further discussion of
the detector background and flat field, grating scatter within the
plane of dispersion, and the use of spectra obtained through the
off-axis aperture to estimate the contribution of diffuse airglow
emission.

\begin{table*}[b]
\label{file_format}
\begin{center}
\small
{\sc Table 3\\
BEFS File Formats}
\vskip 2pt
\begin{tabular}{lcl}
\tableline
\tableline
Field Name & Units & Description \\
\tableline
\multicolumn{3}{c}{First-Order Spectra} \\
\tableline
WAVELENGTH & \AA\ & Wavelength array \\
FLUX\_PHOT & photons cm$^{-2}$ s$^{-1}$ \AA$^{-1}$ & Calibrated flux in photon units \\
FLUX\_ERGS & erg cm$^{-2}$ s$^{-1}$ \AA$^{-1}$ & Calibrated flux in energy units \\
FRAC\_ERR & \nodata & Fractional uncertainty in the flux (1 $\sigma$)\\
X & pixels & Pixel number \\
SPEC & counts & Extracted spectrum \\
BG0 & counts & Spectrum of background region 0 \\
BG1 & counts & Spectrum of background region 1 \\
BACKGROUND & counts & Normalized background, smoothed by 13 pixels \\
BG\_ERR & counts & Uncertainty in the background (1 $\sigma$) \\
DELTAW & \AA\ & Wavelength increment \\
EFF\_AREA & cm$^{-2}$ & Effective area \\
\tableline
\multicolumn{3}{c}{Second-Order Spectra} \\
\tableline
WAVELENGTH & \AA\ & Wavelength array \\
2ND\_COUNTS & counts & Estimated second-order spectrum \\
2ND\_SIGMA & counts & Uncertainty in second-order spectrum (1 $\sigma$) \\
2ND\_AEFF & cm$^{-2}$ & Second-order effective area \\
\end{tabular}
\end{center}
\end{table*}

\section{THE DATA SET}
\label{format}

BEFS data from both the {\it ORFEUS-SPAS~I\/ {\rm and} II}\/ missions
are available from the Multimission Archive at STScI (MAST;
\url{http://archive.stsci.edu/mast.html}).  All files are in FITS
format, with the data stored as one or more binary-table extensions.
The data-reduction procedures employed to produce these files may be
summarized as follows:

1. Accumulate data between designated start and stop times.  Identify
``good times.''  Produce a photon-event file, which lists position and
arrival time {\it (x, y, t)}\/ for each event on a given detector,
including stim pins.  Only events which arrive during good times are
included in the photon-event file.  This step is applied independently
to the data stream for each observation and each detector.

2. Extract photon events from the desired aperture, its associated
background regions, and the nearest stim pin.  Extraction windows are
predefined and are constant throughout each mission, except as
discussed in Section \ref{extraction}.

3. Determine detector dead time from stim-pin counts and exposure time.

4. Correct the wavelength scale for temporal effects (a stretch and shift).

5. Convert the photon-event list into a raw-counts spectrum.  Combine
the background strips and scale to produce a background spectrum.
Smooth the background array by 13 pixels ($\sim$ 0.18 \AA).  Generate
error arrays (assuming Gaussian statistics) for both the target and
background spectra.

6. Apply the flux calibration as follows: \texttt{FLUX\_PHOT = (SPEC -
BKGD) $\times$ DT\_CORR / (EFF\_AREA $\times$ EXPTIME $\times$
DELTA\_WAV)}, where \texttt {FLUX\_PHOT} is the flux-calibrated
spectrum in units of photons cm$^{-2}$ s$^{-1}$ \AA$^{-1}$, \texttt
{SPEC} and \texttt{BKGD} are the raw-counts and background arrays,
\texttt{DT\_CORR} is the dead-time correction, \texttt{EFF\_AREA} the
effective area in cm$^2$, \texttt{EXPTIME} the exposure time in s, and
\texttt{DELTA\_WAV} the width of a pixel in \AA ngstroms.

7. Write out the fully-calibrated spectrum as a FITS binary table
extension according to the format presented in Table 3.

Steps 2--7 are performed independently for each aperture.  We do not
attempt to correct BEFS spectra for detector flat-field effects,
photometric losses due to mis-centering of the target in the aperture,
or resolution degradation resulting from pointing jitter.  We provide,
but do not subtract, FUV airglow (D2) spectra obtained on the 1996
mission.  (Because the D1 and D2 spectra have different point-spread
functions, one cannot simply be subtracted from the other.)

8. Model the second-order contribution to the FUV spectrum and write it
to a second binary table extension in the SPD1 file, using the format
given in Table 3.  The units of the second-order
spectrum are raw counts; gaps in the data are set to zero.  To apply
the correction, smooth the second-order array and subtract it from the
first-order raw-counts spectrum.  Re-calibrate the first-order spectrum
using the recipe given in Step 6.  Photometric uncertainties or
spectral anomalies in the second-order spectra (especially for the {\it
ORFEUS-SPAS~II}\/ data) may require manual intervention.  This step
is performed only on data sets for which EUV spectra are available.

The complete set of software used to extract and calibrate BEFS spectra
(written in C), together with all supporting parameter and calibration
files, are available from MAST.  In addition, we have written two IDL
routines to manipulate BEFS data.  The first, makeair.pro, reads an
airglow (D2) file and scales a synthetic airglow spectrum to the
observed line intensities.  The second, extract\_tt.pro (also available
in C), reads a photon-event file (LST1) and returns the detector X
coordinate and arrival time of each photon falling within a
user-defined region of the detector.  General IDL routines to read and
manipulate BEFS data files are available from the IDL Astronomy User's
Library (\url{http://idlastro.gsfc.nasa.gov}).

\section{DISCUSSION}

\begin{figure}
\epsscale{0.8}
\plotone{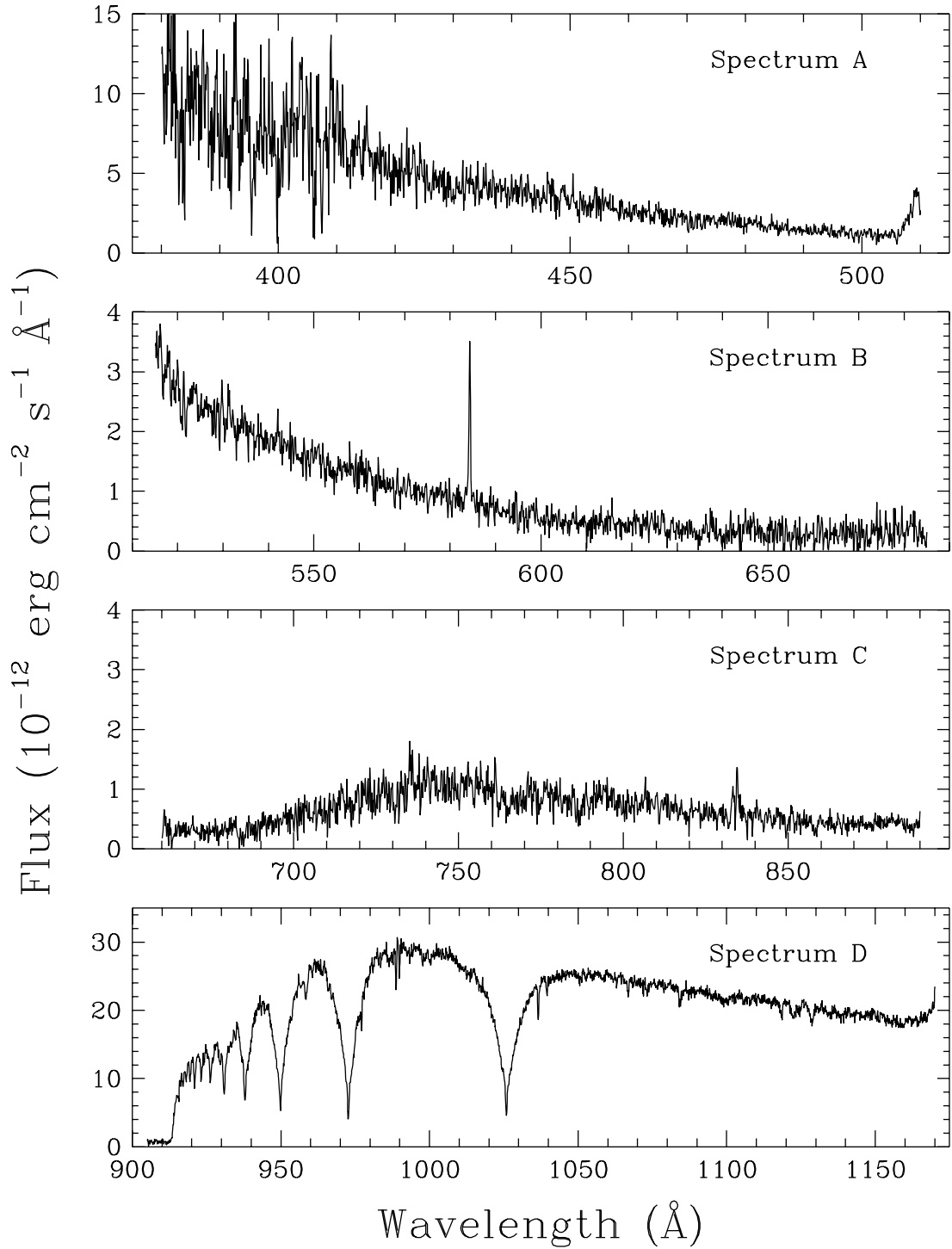}
\figcaption{Complete spectrum of G~191--B2B obtained by the BEFS on the
{\it ORFEUS-SPAS~I}\/ mission.  Spectra A, B, and C are binned by 16
pixels, spectrum D by 8.  Note the bright corner, which contaminates
spectrum A at wavelengths shorter than about 420 \AA, and the
geocoronal \ion{He}{1} $\lambda 584$ and \ion{O}{2} $\lambda \lambda 832.7$--834.5
features in spectra B and C, respectively.  Most of the flux in
spectrum C is second-order light from EUV wavelengths.
\label{g191b2b}
}
\end{figure}

The various instrumental and detector effects discussed above are best
appreciated by an examination of the data.  Figure \ref{g191b2b}
presents the complete spectrum of G~191--B2B obtained on the {\it
ORFEUS-SPAS~I}\/ flight of BEFS.  It is the sum of four individual
pointings with a total integration time of 8100 s.  The individual
spectra from Gratings B and D, which have well-defined spectral
features, were cross-correlated before being combined.

\begin{figure}
\epsscale{0.7}
\plotone{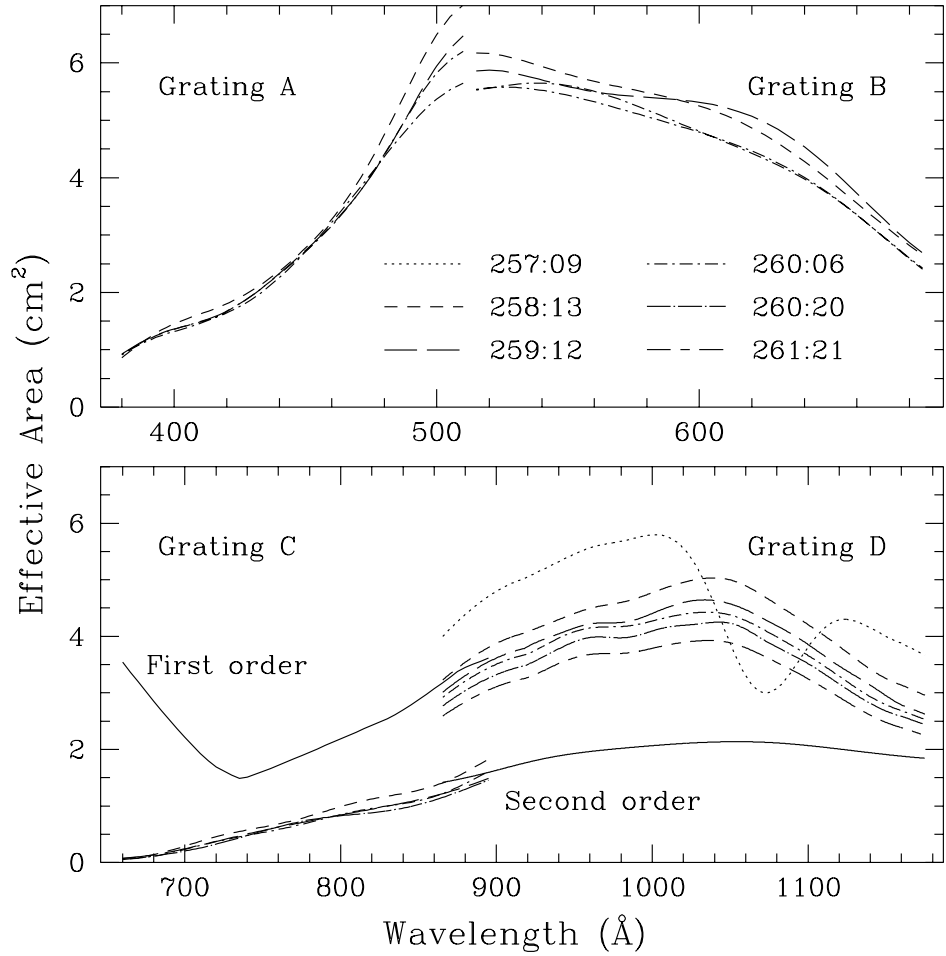}
\figcaption{Effective-area curves for the BEFS on the {\it
ORFEUS-SPAS~I}\/ mission, derived from observations of white dwarf
stars, primarily G~191--B2B.  Differences among the Grating A and D
curves are thought to be real and reflect a decrease in sensitivity
with time.  Differences among the Grating B curves are not significant
given the S/N ratios of the individual spectra; an average of the
Grating B curves is applied to all spectra.  The Grating C spectrum of
G~191--B2B shows no first-order astrophysical flux; the curve presented
is theoretical.  The Grating D effective area shows a slow decrease in
sensitivity with time; we interpolate among these curves to calibrate
our data.  Theoretical second-order curves are scaled to match the flux
seen in spectrum C and in spectrum D below 912 \AA.
\label{o1_flux}
}
\end{figure}

Note the low signal-to-noise ratio at the shortest wavelengths, where
the strong background flux of the bright corner has been subtracted
from spectrum A.  The \ion{He}{1} edge at 504 \AA\ is just visible at
the long-wavelength end of this spectrum.  Spectrum B shows a steep
drop in the continuum flux toward long wavelengths, punctuated by the
geocoronal \ion{He}{1} $\lambda 584$ line.  Because the interstellar
medium toward G~191--B2B is nearly opaque at wavelengths between
650 and 912 \AA, spectrum C is dominated by second-order EUV emission.
The apparent peak at $\sim 750$ \AA\ reflects a dip in the first-order
flux-calibration curve (\fig{o1_flux}) applied by our calibration
software.  The geocoronal \ion{O}{2} $\lambda \lambda 832.8$--834.5
multiplet is weak in this spectrum.  Flux from the bright corner has
been subtracted from the short-wavelength end of spectrum D, but
does not significantly affect the signal-to-noise ratio of this bright
FUV source.  The BEFS FUV spectra of G~191--B2B are analyzed by 
\citet{Hurwitz:95} and \citet{Vennes:96}.

\begin{figure}
\epsscale{0.8}
\plotone{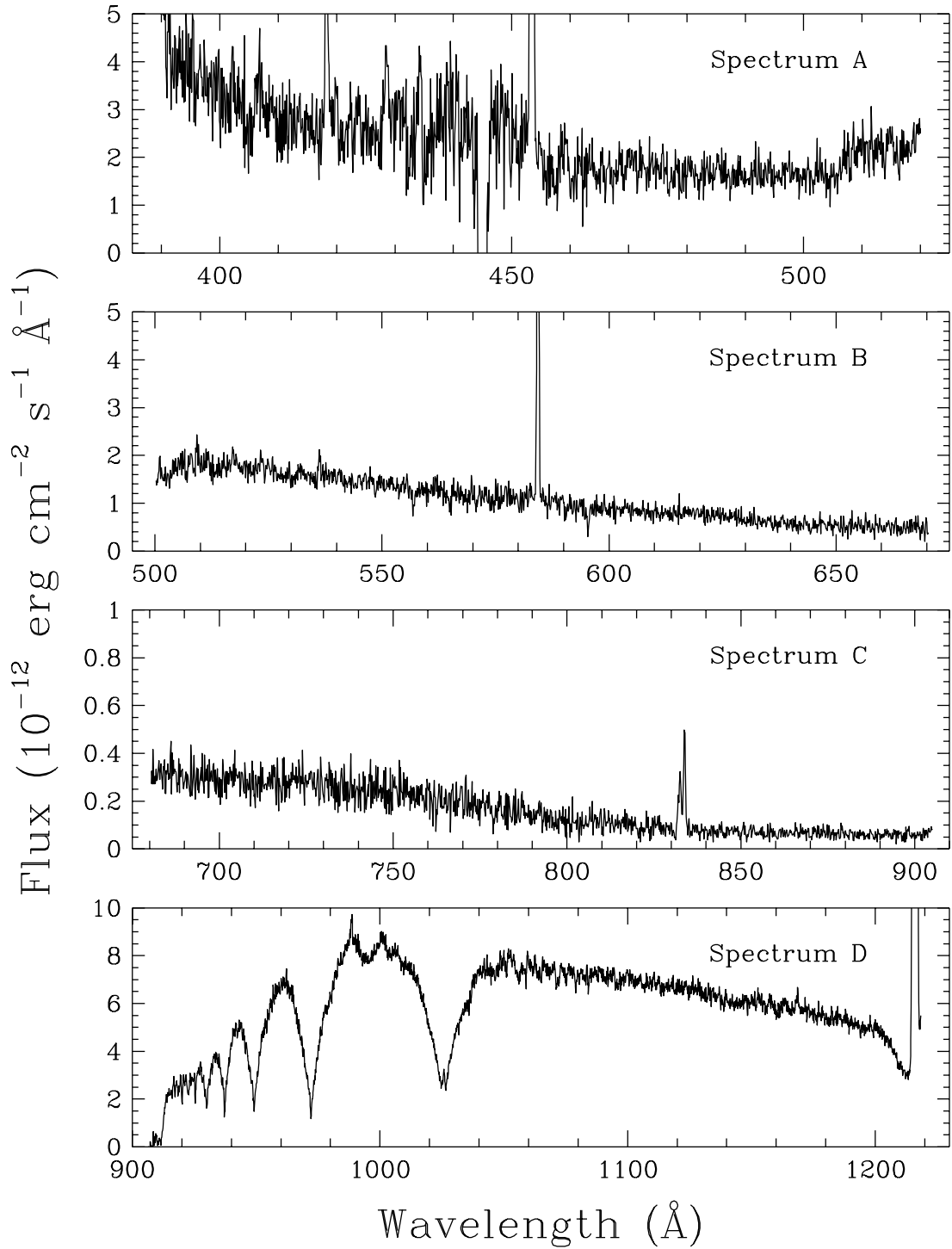}
\figcaption{Complete spectrum of HZ~43 obtained by the BEFS on the {\it
ORFEUS-SPAS~II}\/ mission.  Spectra A, B, and C are binned by 16
pixels, spectrum D by 8.  Because the spectra from Gratings A and B
overlap in this data set, spectrum A is contaminated between about 400
and 460 \AA\ (see discussion in text).  The neutral-hydrogen column
density toward HZ~43 is lower than that toward G~191--B2B, allowing
first-order flux to reach us at the short-wavelength end of spectrum C.
\label{hz43}
}
\end{figure}

Figure \ref{hz43} shows the spectrum of HZ~43 obtained on the {\it
ORFEUS-SPAS~II}\/ mission.  Six individual observations, totaling 8700
s, were combined to make this plot.  Because some of the EUV spectra in
this data set are non-photometric, the individual flux-calibrated
spectra were weighted by their continuum flux, integrated over a
selected wavelength region, before being summed.  The final spectrum
was then rescaled to the intensity of the initial HZ~43 observation,
which for spectra A and B is significantly brighter than subsequent
observations.  The time-dependent stretch and shift in the BEFS
wavelength scale is particularly apparent in these data, which
span the entire two-week mission.  Individual spectra were
cross-correlated before being combined.  Spectrum D was
cross-correlated twice, once using only the Lyman $\gamma$ $\lambda
973$ line, and again using only the short-wavelength wing of the Lyman
$\alpha$ $\lambda 1216$ line.  The resulting spectra were spliced
together at 1050 \AA.  For scientific analysis, a better approach
would be to re-sample all six spectra onto a uniform wavelength grid.

\begin{figure}
\epsscale{1.0}
\plotone{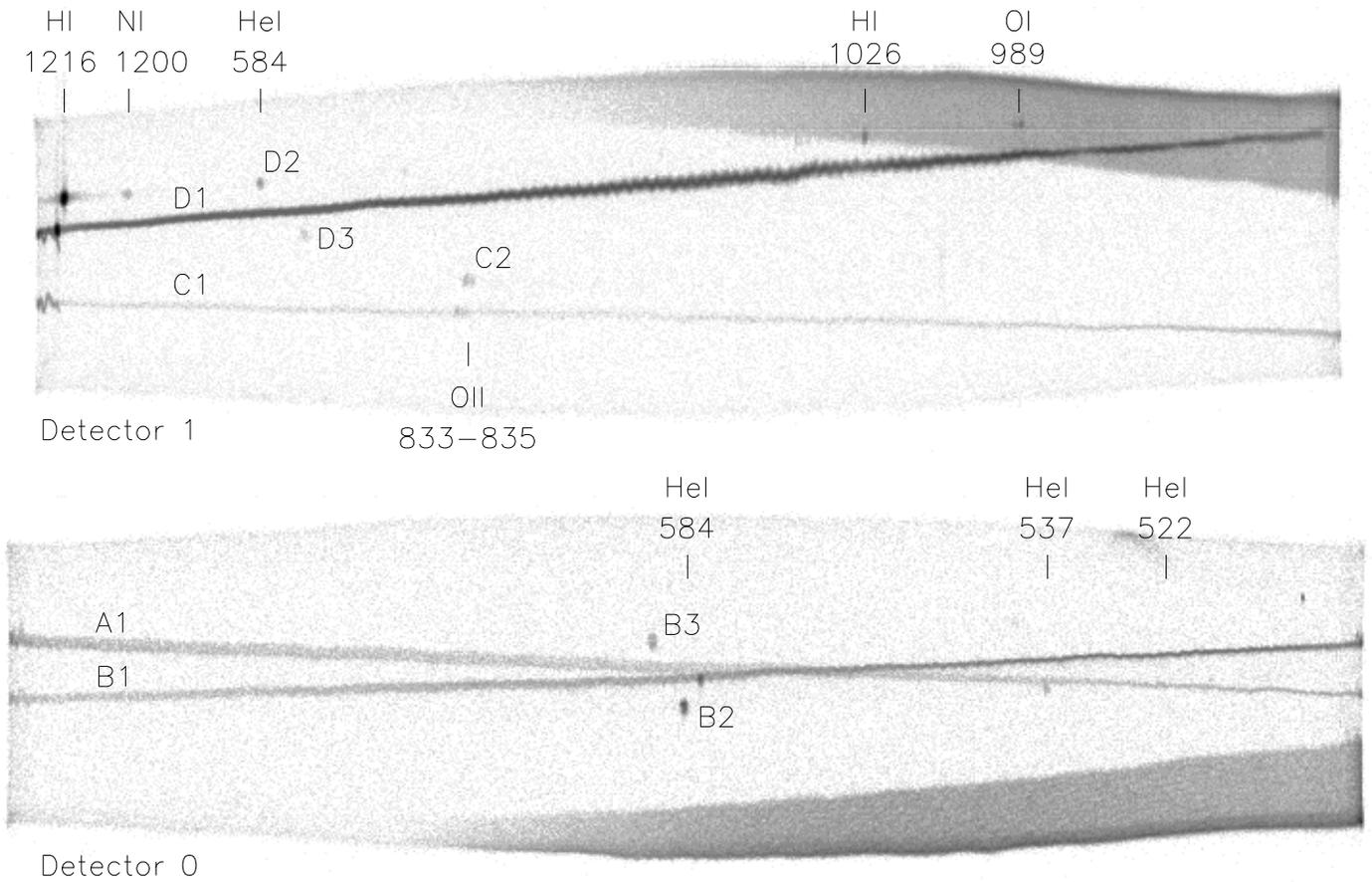}
\figcaption{
Logarithmically-scaled image of Detectors~1 (top) and 0 (bottom) for an
observation of HZ~43 obtained on the {\it ORFEUS-SPAS~II}\/ mission.
The continuum spectra, A1, B1, C1, and D1, are labeled, as
are individual geocoronal lines in the off-axis B2, B3, C2, D2, and D3
spectra.  (The B2 and D3 spectra were never extracted.) Selected
emission features are labeled; note that wavelength increases to the
left.  The Grating A and B spectra cross on Detector~0.  Note the
bright corners on both detectors.
\label{image}
}
\end{figure}

Spectrum A is no longer contaminated by the bright corner, but its
overlap with the Grating B spectra severely complicates the region
between 400 and 460 \AA.  Our software attempts to model and subtract
this contamination, but is not entirely successful; for example, the
strong emission features seen in this region are actually geocoronal
\ion{He}{1} lines from spectra B1 and B2 (see \fig{image}).  The
\ion{He}{1} edge at 504 \AA\ is present in both spectra A and B, and
the lower neutral-hydrogen column density toward HZ~43 allows
first-order flux to reach us at the short-wavelength end of spectrum
C.  The bright FUV spectrum of HZ~43 is relatively immune to the bright
corner, but its high signal-to-noise ratio highlights a number of
flat-field effects, as discussed in Section \ref{flat_field_text}.
Also apparent are the diffuse emission features of \ion{O}{1} $\lambda
989$, Lyman $\beta$ $\lambda 1026$, and \ion{He}{1} $\lambda 584$ in
second order.  The BEFS FUV observations of HZ~43 are analyzed by
\citet{Dupuis:98}.

BEFS, the Berkeley Extreme and Far-Ultraviolet Spectrometer, offers a
unique combination of spectral resolution and effective area in the
comparatively unexplored FUV wavelength band.  We are pleased to make
the entire BEFS data set available to the community in a standard
(FITS) format with improved flux and wavelength calibrations.

\acknowledgments

\begin{figure}
\epsscale{0.8}
\plotone{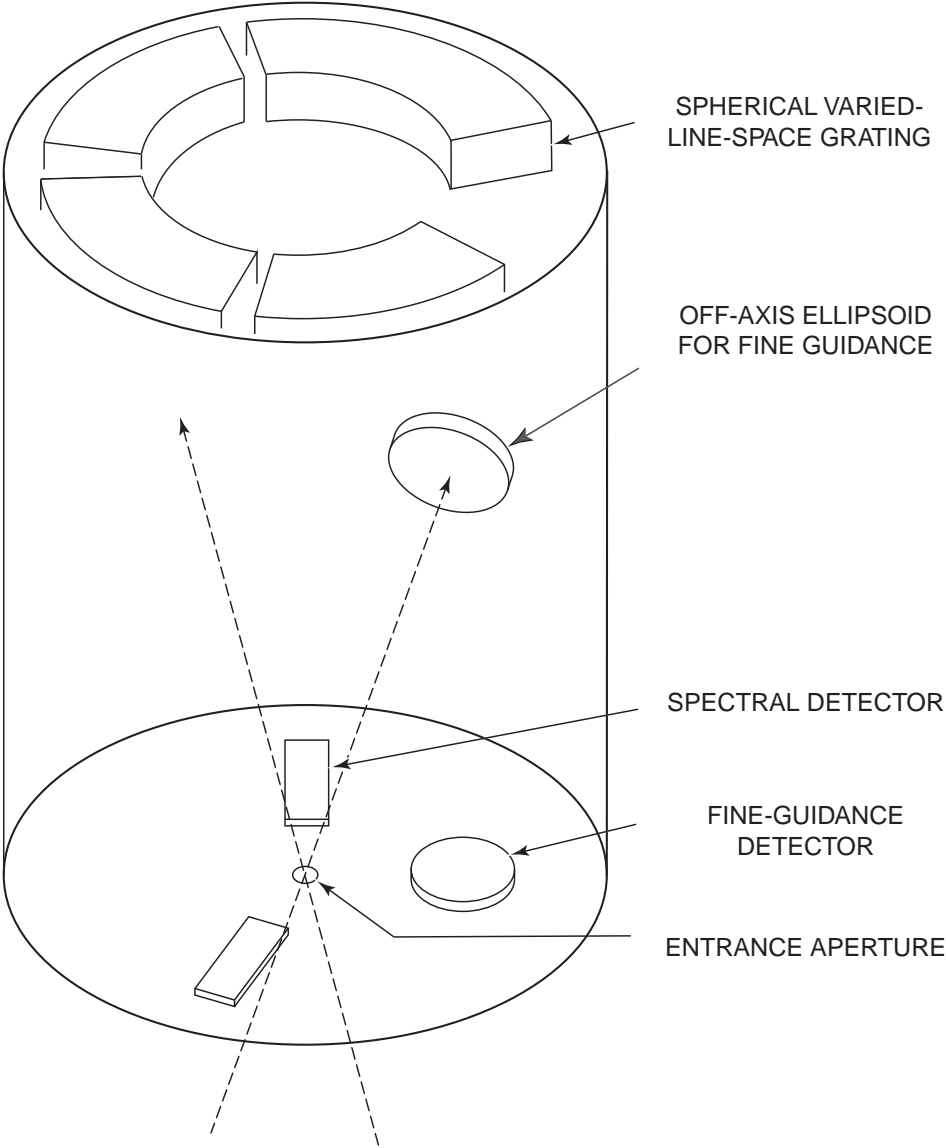}
\figcaption{
Schematic layout of the BEFS.  The spectrograph entrance aperture
lies at the focal point of the telescope.
\label{schematic}
}
\end{figure}

We thank S.~Vennes for providing his grid of synthetic white-dwarf
spectra, B.~Espey for his assistance with the wavelength calibration,
and V. Chauvet for drawing \fig{schematic}.  Thanks to the staff of
MAST for their help and advice in preparing our data for submission to
the archive.  We acknowledge our colleagues on the \orf\ team and the
many NASA and DARA personnel who helped make the {\it ORFEUS-SPAS}\/
missions successful.  This work is supported by NASA grant NAG5-696.




\clearpage

\end{document}